\documentclass[conference]{IEEEtran}
\IEEEoverridecommandlockouts
\usepackage{cite}
\usepackage{amsmath,amssymb,amsfonts}
\usepackage{algorithmic}
\usepackage{graphicx}
\usepackage{textcomp}
\usepackage{xcolor}
\usepackage{hyperref}
\usepackage{orcidlink}
\usepackage{notes2bib}

\def\BibTeX{{\rm B\kern-.05em{\sc i\kern-.025em b}\kern-.08em
    T\kern-.1667em\lower.7ex\hbox{E}\kern-.125emX}}
\begin{document}


\title{Numerical limits in the integration of Vlasov-Poisson equation for Cold Dark Matter\\
\thanks{
This paper is supported by the  Fondazione ICSC, Spoke 10 "Quantum Computing" and Spoke 3 "Astrophysics and Cosmos Observations". National Recovery and Resilience Plan (Piano Nazionale di Ripresa e Resilienza, PNRR) Project ID CN-00000013 "Italian Research Center on  High-Performance Computing, Big Data and Quantum Computing"  funded by MUR Missione 4 Componente 2 Investimento 1.4: Potenziamento strutture di ricerca e creazione di "campioni nazionali di R$\&$S (M4C2-19 )" - Next Generation EU (NGEU)

}
}

\author{\IEEEauthorblockN{1\textsuperscript{st} Luca Cappelli}
\IEEEauthorblockA{\textit{Dipartimento di Fisica Universita' di Trieste} \\
\textit{INAF-OATs} \\
Trieste, Italy \\
\orcidlink{https://orcid.org/0009-0009-1169-8380}
}
\and
\IEEEauthorblockN{2\textsuperscript{nd} Giuseppe Murante}
\IEEEauthorblockA{\textit{INAF-OATs} \\
Trieste, Italy \\}
\and
\IEEEauthorblockN{3\textsuperscript{rd} Stefano Borgani}
\IEEEauthorblockA{\textit{Dipartimento di Fisica Universita' di Trieste} \\
\textit{INAF-OATs} \\
Trieste, Italy \\
\orcidlink{https://orcid.org/0000-0001-6151-6439}}
\and
\IEEEauthorblockN{4\textsuperscript{rd} Andrea Bulgarelli}
\IEEEauthorblockA{\textit{INAF-OAs } \\
Bologna, Italy \\
}
\and
\IEEEauthorblockN{6\textsuperscript{rd} Nicol\'o Parmiggiani}
\IEEEauthorblockA{\textit{INAF-OAs } \\
Bologna, Italy \\
}
\and
\IEEEauthorblockN{7\textsuperscript{rd} Massimo Meneghetti}
\IEEEauthorblockA{\textit{INAF-OAs } \\
Bologna, Italy \\
}
\and
\IEEEauthorblockN{8\textsuperscript{rd} Carlo Burigana}
\IEEEauthorblockA{\textit{INAF-IRA} \\
Bologna, Italy \\
}
\and
\IEEEauthorblockN{9\textsuperscript{rd} Tiziana Trombetti}
\IEEEauthorblockA{\textit{INAF-IRA} \\
, Italy \\
}
\and
\IEEEauthorblockN{10\textsuperscript{rd} Giuseppe Sarracino }
\IEEEauthorblockA{\textit{INAF-O.A. di Capodimonte} \\
Napoli, Italy \\
}
\and
\IEEEauthorblockN{11\textsuperscript{rd} Vincenzo Testa }
\IEEEauthorblockA{\textit{INAF-O.A. di Capodimonte} \\
Napoli, Italy \\
}
\and
\IEEEauthorblockN{12\textsuperscript{rd} Farida Farsian}
\IEEEauthorblockA{\textit{NAF-O.A Catania} \\
Catania, Italy \\
}
\and
\IEEEauthorblockN{13\textsuperscript{rd} Alessandro Rizzo}
\IEEEauthorblockA{\textit{NAF-O.A Catania} \\
Catania, Italy \\
}
\and
\IEEEauthorblockN{14\textsuperscript{rd} Francesco Schillir\'o}
\IEEEauthorblockA{\textit{NAF-O.A Catania} \\
Catania, Italy \\
}
\and
\IEEEauthorblockN{15\textsuperscript{rd} Vincenzo Fabrizio Cardone }
\IEEEauthorblockA{\textit{INAF-O.A. Roma} \\
Roma, Italy \\
}
\and
\IEEEauthorblockN{16\textsuperscript{rd} Roberto Scaramella}
\IEEEauthorblockA{\textit{INAF-O.A. Roma} \\
Roma, Italy \\
}
}


\maketitle

\begin{abstract}
The Vlasov-Poisson systems of equations (VP) describes the evolution of a distribution of collisionless particles under the effect of a collective-field potential. VP is at the basis of the study of the gravitational instability of cosmological density perturbations in Dark-Matter (DM), but its range of application extends to other fields, such as  plasma physics. 
    
In the case of Cold Dark Matter, a single velocity is associated with each fluid-element (or particle) , the initial condition presents a stiff discontinuity. This creates problems such as diffusion or negative distribution function when a grid based method is used to solve VP. In this work we want to highlight this problem, focusing on the technical aspects of this phenomenon. By comparing different finite volume methods and a spectral method we observe that, while all integration schemes preserve the invariants of the system (e.g, energy), the physical observable of interest, i.e., the density, is not correctly reproduced. We thus compare the density obtained with the different Eulerian integration schemes with the result obtained from a reference N-body method. We point out that the most suitable method to solve the VP system for a self-gravitating system is a spectral method.
\end{abstract}


\section{Introduction}
    Modern cosmological models, particularly the $\Lambda CDM$ framework, provide the foundation for our current understanding of the universe \cite{Planck2020}. This model suggests that a significant portion of the universe is composed of dark energy, an unknown form of energy responsible for the accelerated expansion, and cold dark matter (CDM), a collisionless, non-baryonic component that plays a critical role in structure formation. While the precise nature of these dark components remains elusive, it is broadly accepted that the gravitational instability of minute CDM density fluctuations, originating in the early universe, drives the formation of cosmic structures. These structures range from galactic scales (kiloparsecs, kpc) to the vast scales of the cosmic web (gigaparsecs, Gpc) \cite{Weinberg2013}.

    Developing an accurate numerical description of density perturbation evolution in the strongly non-linear regime, while avoiding the assumption underlying the N-body treatment, can reveal subtle details on the formation and dynamics of cosmic structures.  When compared to observational data, these predictions can further validate the standard $\Lambda$CDM paradigm, or shed light on the fundamental properties and nature of dark matter in regimes that can not be accessed by N-body techniques~\cite{Mocz_towards, Cappelli}.

    In these regards, the evolution of cold dark matter (CDM) can be effectively modeled as a collisionless fluid, governed by the Vlasov-Poisson equation. This equation describes the time evolution of the phase-space distribution function, thereby capturing the dynamics of the system in a 6-dimensional phase space, plus time, thus adding complexity as the resolution of the problem increases. For this reason, it is often preferable not to directly integrate the Vlasov-Poisson (VP) system but to employ alternative approaches. 
    The most common method involves implementing an N-body scheme (e.g., particle mesh (PM)), which discretizes the phase-space distribution function of the fluid via the Vlasov-Poisson (VP) system of equations \cite{Springel2016}. This approach allows for the numerical tracking of interactions between discrete particles that approximate the behavior of the continuous fluid with the drawback of producing inaccurate results in sparsely populated regions \cite{Yoshikawa_2013} and might require the introduction of a gravitational softening term to avoid artificial large-angle scattering of particles caused by close encounters.  
    
    An alternative to N-body methods is provided by Schrödinger methods, where the Schrödinger-Poisson (SP) equation is solved instead of the VP system \cite{widrow-kaiser, mocz_schrodinger-poissonvlasov-poisson_2018}. In the limit where the dynamic scale of the problem, $\hbar/m$, becomes very small, the VP results can be recovered. This approach proved also to be a good starting point for the development of quantum algorithms for cosmological simulations \cite{Cappelli}, opening the way for possible advantages in the future.

    In spite of the challenges, several alternative grid-based methods have been developed for simulating the dynamics of self-gravitating systems. These include the pioneering splitting method \cite{cheng-knorr}, the more general semi-Lagrangian methods \cite{sonnendrucker}, 
    finite volume methods \cite{filbet-sonnendrucker-fpc}, 
    spectral methods \cite{klimas},
    and more recently, even for cosmological simulations \cite{mocz_schrodinger-poissonvlasov-poisson_2018,oliver-hahn,Yoshikawa_2013}. 
    A comparative analysis of several of these techniques is provided in \cite{comparison, new-method-VP}.

    To facilitate a comparison between the aforementioned approaches -- Particle-Mesh (PM), Schrödinger-Poisson (SP), and Vlasov-Poisson (VP) -- it is essential to integrate a system starting from the same initial conditions, thus the necessity of this work. Here we aim to highlight the importance of choosing the appropriate integration scheme for the Vlasov-Poisson system, particularly in those cases where there is a sharp discontinuity in the initial conditions. We identify a specific set-up in which macroscopic physical quantities (e.g., energy, momentum, $L_1$ norm, etc.) are preserved but the density is not accurately represented. 
    As a matter of fact, suitable solution to this specific kind of problem has not been discussed so far in literature. This represents the main motivation for this work.
    
    We propose here a comparison between a flux-conserving method and a spectral method, considering both "smooth" and  discontinuous initial conditions. 
    We then discuss the reasons behind the misrepresentation of density, even in presence of conservation of other quantities. We finally propose a solution for this limitation, also identifying which method is most suitable for addressing this specific problem.

    In summary, the innovative aspect of this work consists in identifying the origin of numerical challenges arising in a pathological case of velocity discontinuity in the initial phase-space configuration, which could have implications in the context of cosmological applications, thereby proposing a solution for the treatment of such cases.
    
    The paper is organized as follows: Sec.~\ref{sec: vlasov for cdm} introduces the Vlasov-Poisson equation and the quantities that should be conserved during its evolution. Also, is presented a brief explanation of why we are interested in using sharply discontinuous initial conditions. Sec.~\ref{sec: numerical schemes} addresses the numerical approach to the problem, presenting the two integration schemes employed and providing details on timestep selection, discretization, and the practical implementation of the integration process. Sec.~\ref{sec: numerical test} presents the results of the numerical tests conducted under the two different initial conditions. Provides also an analysis and interpretation of the results, describing the nature of the numerical issue and proposing a viable solution.

%
\begin{figure*}[t]
    \centering
    \includegraphics[width = 1.7\columnwidth]{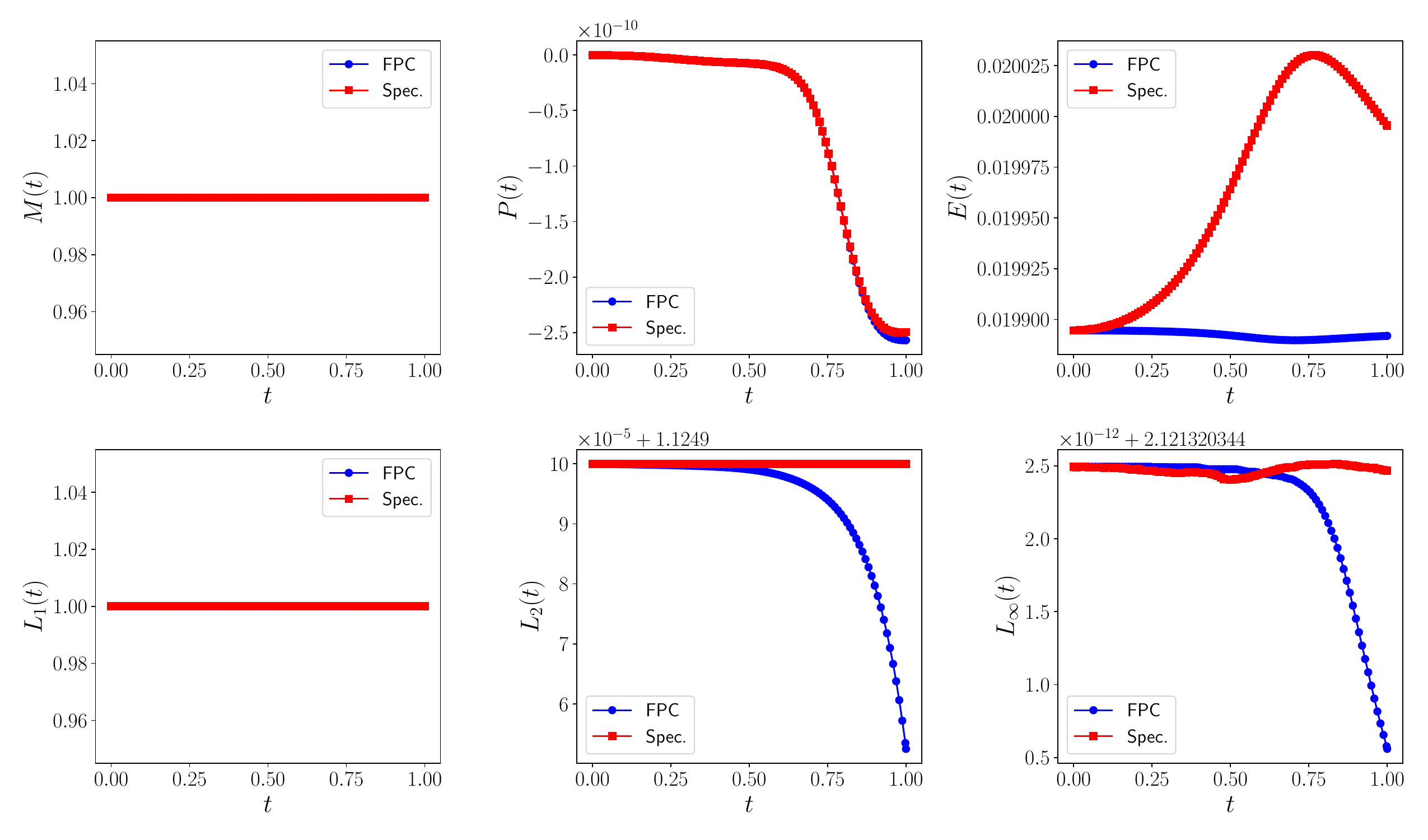}
    \caption{Smooth initial condition : Evolution of physical observables (top row) and norm (bottom row) as a function of time in a simulation of a VP system with $N = 2^9$, $M = 2^{10} + 1$. Starting from the top-left corner we find, Mass $M(t)$, momentum $P(t)$ and energy $E(t)$; from bottom-left corner we find the $L_1$, $L_2$ and $L-\infty$ norm of the distribution function $f(x,v,t)$. }
    \label{fig: Maxwellian observables}
\end{figure*}
%

    \section{Vlasov Equation for CDM}\label{sec: vlasov for cdm}
    The Vlasov equation describes the evolution of a distribution function $f(\boldsymbol{x}, \boldsymbol{v}, t)$ of collisionless matter.
    \begin{equation}
        \label{eq: VP}
        \frac{\partial f(\boldsymbol{x}, \boldsymbol{v}, t)}{\partial_t }
        + \boldsymbol{v} \cdot \frac{\partial f(\boldsymbol{x}, \boldsymbol{v}, t)}{\partial_{\boldsymbol{x}}} 
        - \frac{\partial \phi}{\partial_{\boldsymbol{x}}} \cdot \frac{\partial f(\boldsymbol{x}, \boldsymbol{v}, t)}{\partial_{\boldsymbol{v}}} = 0.
    \end{equation}
    When coupled with a gravitational potential is referred commonly as Vlasov-Poisson (VP) and is used to described the evolution of dark matter perturbations. 
    \begin{equation}
        \label{eq: Poisson pot}
         \nabla^2 \phi(\boldsymbol{x}, t) = 4\pi G \left(\rho(\boldsymbol{x}, t) - \rho^* \right),
    \end{equation}
    where $\rho(\boldsymbol{x},t) = \int f(\boldsymbol{x}, \boldsymbol{v},t) \, d^3 \boldsymbol{v}$ and $\rho^*$ is a reference density.
    
    As anticipated in the introduction, one of the possible way to solve the VP system is mapping it onto the Schr\"odinger equation~\cite{klimas, mocz_schrodinger-poissonvlasov-poisson_2018,Cappelli}. Such Schrödinger method is expected to perform particularly well in the Cold Dark Matter (CDM) scenario, where each density element is associated with a single velocity. To compare how well SP and PM methods reproduce VP results, it is necessary to integrate the VP system. This creates the need to explore cases with sharply discontinuous initial conditions. Specifically, the CDM clause translates into a distribution in the velocity given by a Dirac's delta $\rho(v) = \delta(\boldsymbol{v} - \boldsymbol{v_0})$. 

    
    \subsection{Conservation laws}
    A good numerical integrator should be able to represent the right evolution of the distribution, which is not known a priory. Is it proven however that the Vlasov-Poisson system verifies a certain set of conservation properties. We will study the behavior of such proprieties to verify which numerical integration method is best suited in our scenario. The properties we want to check are the following:
    \begin{itemize}
        \item Maximum principle
            $$
                0 \leq f(\mathbf{x}, \mathbf{v}, t) \leq \max_{(\mathbf{x}, \mathbf{v})}\left\{f_0(\mathbf{x}, \mathbf{v}) \right\}
            $$
        \item Conservation of $L^p$ norm, $p \in \left[ 1 , \infty \right)$
            \begin{equation}\label{eq: Lp norm}
                || f_t ||_{L^p} = \int d\mathbf{x}\, d\mathbf{v} \left[
                    f(\mathbf{x}, \mathbf{v}, t)
                \right]^p \, .
            \end{equation}
        \item Total mass conservation. For any volume $V$ of phase space
            \begin{equation}\label{eq: Volume expression}
            M = \int_V d\mathbf{x}\, d\mathbf{v} f(\mathbf{x}, \mathbf{v}, t) \, .
            \end{equation}
        \item Momentum conservation 
        \begin{equation}
            \label{eq: momentum expression}
            P = \int_V d\mathbf{x}\, d\mathbf{v} f(\mathbf{x}, \mathbf{v}, t) \mathbf{v} \, .
        \end{equation}
        \item Energy conservation
            \begin{equation}
                \label{eq: energy}
                E = \frac{1}{2}\int_V d\mathbf{x}\, d\mathbf{v} f(\mathbf{x}, \mathbf{v}, t) \left( \mathbf{v}^2 + \phi(\boldsymbol{x},t) \right)  \, ,
            \end{equation}
            with $\phi(\mathbf{x},t)$ from Eq.~\eqref{eq: Poisson pot}.
    \end{itemize}


\begin{figure*}[t]
    \centering
    \includegraphics[width = 1.7\columnwidth]{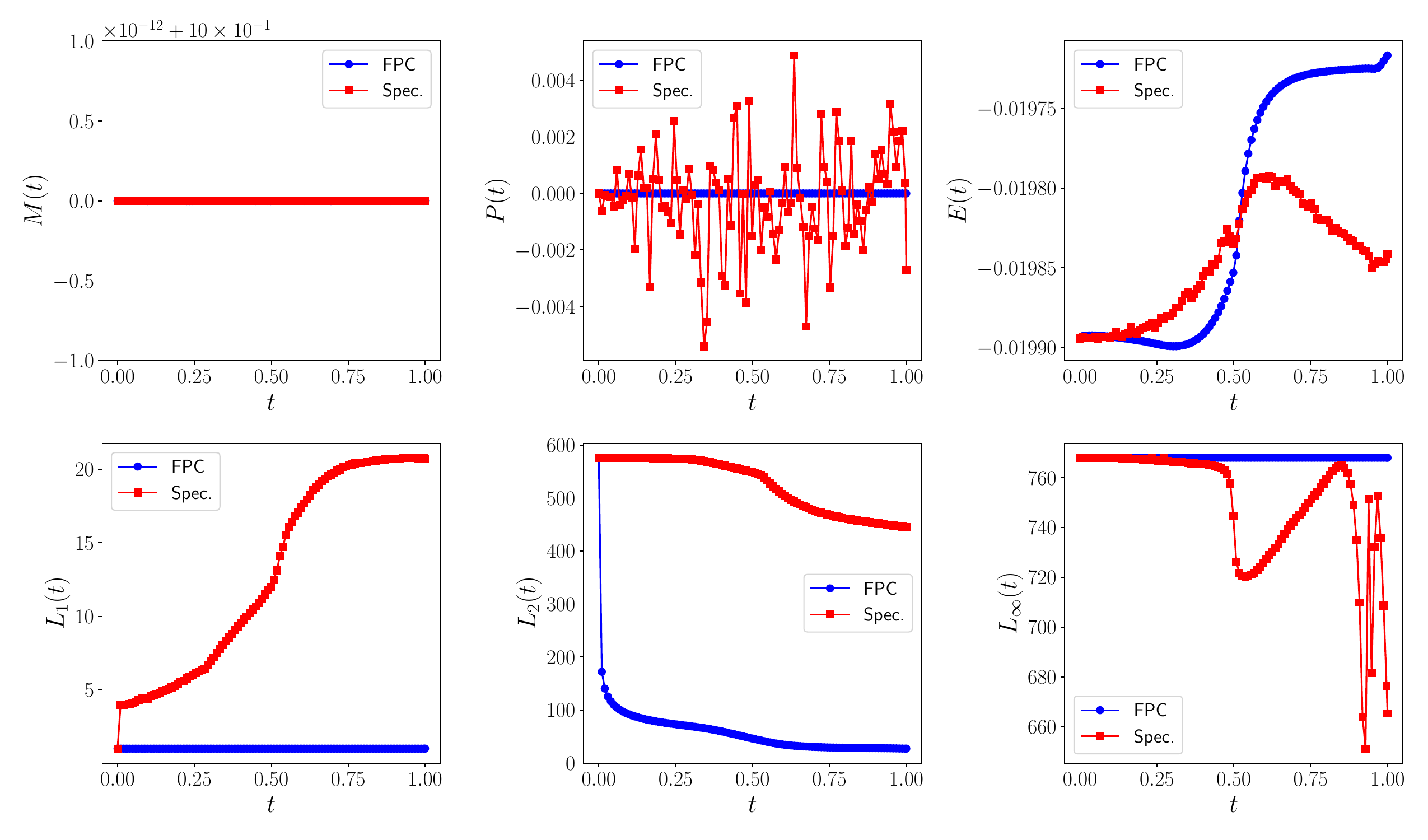}
    \caption{CDM stiff initial condition : Evolution of physical observables (top row) and norm (bottom row) as a function of time in a simulation of a VP system with $N = 2^{10}$, $M = 2^{11} + 1$. Starting from the top-left corner we find, Mass $M(t)$, momentum $P(t)$ and energy $E(t)$; from bottom-left corner we find the $L_1$, $L_2$ and $L-\infty$ norm of the distribution function $f(x,v,t)$. }
    \label{fig: CDM delta observables}
\end{figure*}


    \section{Numerical Schemes}\label{sec: numerical schemes}
    In this work, we will focus exclusively on the one-dimensional case, but the analysis can be easily extended to higher dimensions. 
    
    To integrate Eqs.~\eqref{eq: VP} and \eqref{eq: Poisson pot}, we adopt the splitting scheme~\cite{cheng-knorr} dividing VP into two advection equations, one for the velocity and one for the position
    \begin{align}
        \label{eq: advection x}
        \frac{\partial f(x,v,t)}{\partial t} + v \, \frac{\partial f(x,v,t)}{\partial x} = 0 \,, 
        \\
        \label{eq: advection v}
        \frac{\partial f(x,v,t)}{\partial t} - \frac{\partial \phi}{\partial x} \, \frac{\partial f(x,v,t)}{\partial v} = 0   \, .
    \end{align}
    We then proceed with the discretization. Following a grid-based approach, the spatial domain of length $L$ is divided into $N$ grid points with periodic boundary conditions, while the velocity domain is restricted to the dynamic range $[-V_{\text{max}}, V_{\text{max}}]$ and discretized into $M$ grid points with vanishing boundary conditions, ensuring that matter elements are excluded when their velocity exceeds $|v| > V_{\text{max}}$.

     The discrete representation of $f(x,v,t)$ takes now the form of an $M \times N$ matrix. Consequently, Eq.~\eqref{eq: advection x} gives us $M$ advection equations for the position, with each vector consisting of $N$ components
    \begin{equation}
        \label{eq: discretized adv. x}
        \frac{\partial f}{\partial t}(x_i,v_j, t) = - v_j \, \frac{\partial f}{\partial x}(x_i, v_j,t) \, ,
    \end{equation}
    and Eq.~\eqref{eq: advection v}, $N$ equations of $M$ components for the velocity
    \begin{equation}
        \label{eq: discretized adv v}
        \frac{\partial f}{\partial t}(x_i,v_j, t) =  \frac{\partial \phi}{\partial x}(x_i) \, \frac{\partial f}{\partial v}(x_i, v_j,t) \, ,
    \end{equation}
    To go from $f^n$ to $f^{n+1}$ we employ a leapfrog scheme, bringing the accuracy to $\mathcal{O}(\Delta t^2)$; by solving Eq.~\eqref{eq: advection v} for half time step
    \begin{equation}
        \label{eq: leapfrog scheme 1}
        f^*(x,v) = f^n(x, v - F \Delta t / 2) \,
    \end{equation}
    then solving Eq.~\eqref{eq: advection x} for a full time step
    \begin{equation}
        \label{eq: leapfrog scheme 1}
        f^{**}(x,v) = f^*(x - v \Delta, v) \, .
    \end{equation}
    With $f^{**}$ we solve the Poisson equation~\eqref{eq: Poisson pot} and find a new $F = -\partial_x \phi[f^{**}]$. As final step, we solve Eq.~\eqref{eq: advection v} for a second half time step. 
    \begin{equation}
        \label{eq: leapfrog scheme 1}
        f^{n+1}(x,v) = f^{**}(x, v - F \Delta t / 2) \,
    \end{equation}
    \subsection{Flux positive conservative method (FPC)}
    Among the various grid methods proposed for the solution of VP, the FPC approach~\cite{filbet-sonnendrucker-fpc} has proven to behave considerably well in different scenarios, preserving positivity of the distribution function and the maximum principle. We have chosen to display only the results obtained with the FPC method based on the results presented in~\cite{comparison} and the fact that it has proven to be the better choice within the pool of semi-lagrangian method we tested from~\cite{new-method-VP}. We will briefly summarize the FPC approach for a single advection equation. 
    
    By writing the value $f_i = f(x_i, v,t)$ as the mean value on the cell we have 
    \begin{equation}
        \label{eq: f_i formula PFC}
        f_i^n \Delta x = \int_{x_i-\Delta x/2}^{x_i+\Delta x/2} f(x,v,t^n) \,  dx \, ,
    \end{equation}
    and 
    \begin{equation}
        \label{eq: fluxes PFC}
        \Phi_{\pm} = \int_{X(t^n, \, t^{n+1}, \, x \pm \Delta x/2)}^{x \pm \Delta x/2} f(x,v,t^n) \, dx \, .
    \end{equation}
    Here with $X(t^n, \, t^{n+1}, \, x \pm \Delta x/2)$ we refer to the value of the variable $x$ if we trace back the solution from the point $(x_i\pm \Delta x / 2, t^{n+1}
    )$. The interpolation of the function is done by primitive reconstruction and the positivity is guarantied by the presence of regulators. Using Eqs.~\eqref{eq: f_i formula PFC},~\eqref{eq: fluxes PFC} we can find the expression for the distribution function at the next timestep
    \begin{equation}
        f_i^{n+1} = f_i + \frac{1}{\Delta x} \left( \Phi_- - \Phi_+ \right) \, .
    \end{equation}
    \subsection{Spectral method}
    The spectral method proposed by~\cite{klimas} solves the one-dimensional Vlasov–Poisson and Vlasov–Maxwell systems using a spectral approach. The distribution function \( f_N(t, x, v) \) is approximated by a partial sum of a Fourier series:
    \begin{equation}
    \label{eq: 1 spectra;}
    f_N(t, x, v) = \sum_{k=-N}^N \hat{f}_k(t, v) \exp\left(-i \frac{2\pi k x}{L}\right),
    \end{equation}
    where \( L \) is the domain length in the \( x \)-direction, and \( \hat{f}_k(t, v) \) are the Fourier coefficients. These coefficients are computed from the discrete values of \( f \) sampled at grid points \( x_j \) as:
    \begin{equation}
    \label{eq: 2spectral}
    \hat{f}_k(t, v) = \sum_{j} f(t, x_j, v) \exp\left(i \frac{2\pi k x_j}{L}\right), \quad k \in \{-N, \dots, N\}.
    \end{equation}
    By means of Eqs.~\eqref{eq: 1 spectra;},~\eqref{eq: 2spectral} the evaluation of the advection term in Eq.~\eqref{eq: leapfrog scheme 1} can be treated a phase shift applied in Fourier space. For each mode \( k \), this shift is expressed as:
    \begin{equation}
    \label{eq: spctral shift}
    \hat{f}^*_k(v) = \hat{f}_k(t_n, v) \exp\left(-i \frac{2\pi k}{2L}v \Delta t\right),
    \end{equation}
    where \(\Delta t\) is the time step. Thus the algorithm proceeds via a split-step scheme. First, a forward Fast Fourier Transform (FFT) is performed to compute the Fourier coefficients \(\hat{f}_k\) from the spatial distribution \( f(x, v) \). Next, the phase shift is applied to evolve the solution in Fourier space. Finally, an inverse FFT is used to reconstruct the updated distribution function in physical space.

    \subsection{Poisson Solver}\label{subsec : solution of poisson}
    The Poisson equation we solve in the 1D scenario is a slight rearrangement of Eq.~\eqref{eq: Poisson pot}
    \begin{equation}\label{eq: Nuova Poisson}
        \Delta \phi(x) = \alpha \left( \frac{\rho}{\rho^*} - 1\right) \, ,
    \end{equation}
    where $\alpha = 4\pi G \rho^*$. To numerically solve such equation we employ the convolution approach using the Fourier transform from~\cite{hockney-eastwood}. 

    Under periodic boundary conditions, the discrete Fourier transform (DFT) of the density, $\hat{\rho}(k)$, is calculated using the fast Fourier transform (FFT). The corresponding gravitational potential in Fourier space is expressed as:
    \begin{equation} \label{eq : potential in fourier}
    \hat{\phi}(k) = \hat{G}(k) \hat{\rho}(k),
    \end{equation}
    where $\hat{G}(k)$ is the Green's function in Fourier space
    \begin{equation}
        \hat{G}(k) = - \frac{\alpha \Delta x^2}{4\sin^2{(k \Delta x / 2)}} \, .
    \end{equation}

\begin{figure}[tp]
    \includegraphics[width = .7 \columnwidth]{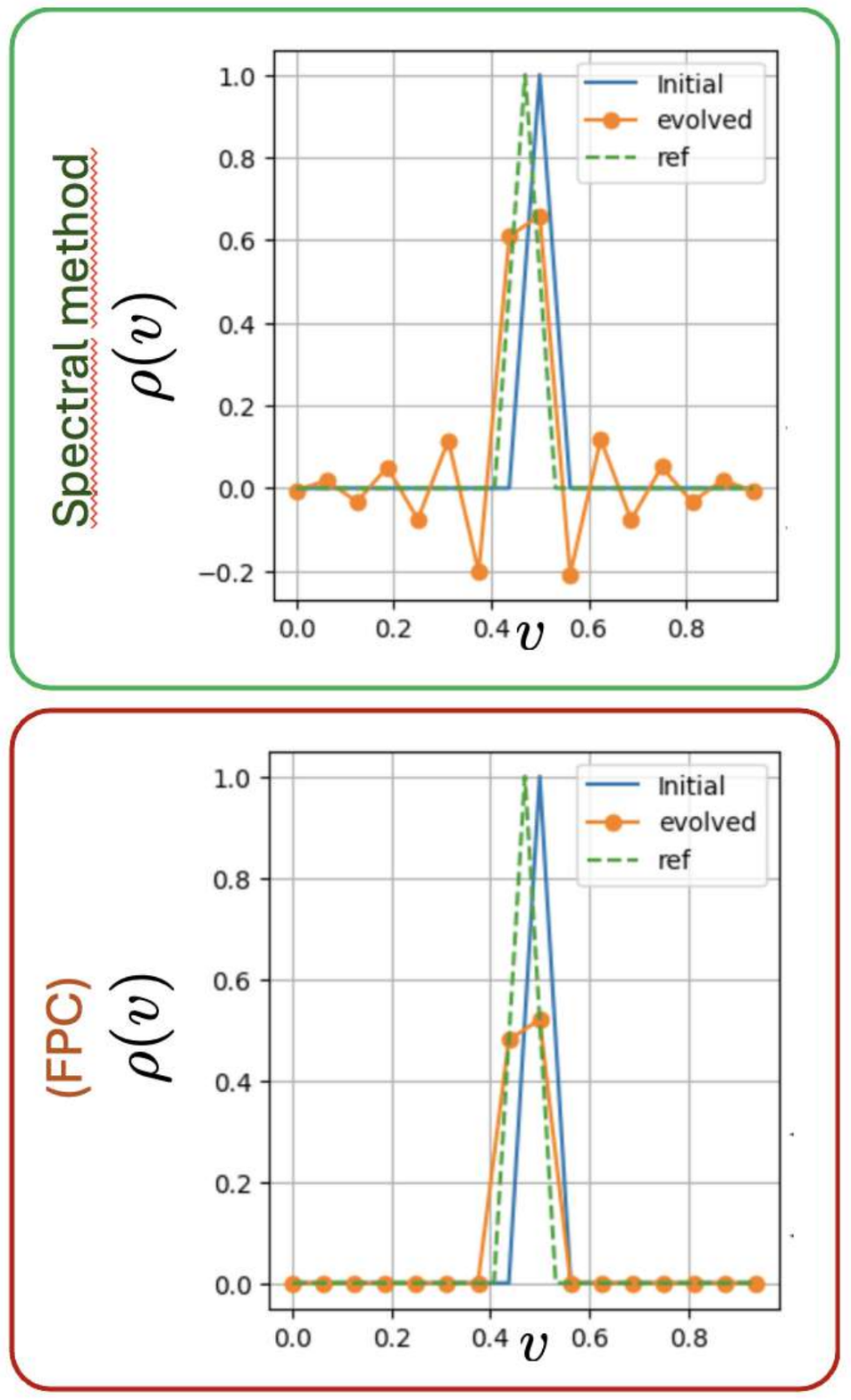}
    \centering
    \caption{Toy model of the first iteration of FPC and spectral method on a Dirac's delta. The blue line represents the initial condition, the green dashed line the analytical expression and the yellow points represents the evolution that is done by the integration scheme.}
    \label{fig: toy_model}
\end{figure}

    \section{Numerical implementation}
    The test cases that we considered for this work have small size, and were only conceived for the purpose of comparing the accuracy of three approaches, rather than their computational performances. Given the size of the problem treated, a proper parallelization scheme is not strictly necessary; all simulations were run on a laptop with typical runtimes always shorter than two hours.

    The code is written in Python, utilizing \texttt{numpy} and \texttt{scipy} to implement the FPC scheme. A significant portion of the code was vectorized using \texttt{numpy}, which improved runtime performance. For the spectral method, we employed the \texttt{fft} library from \texttt{scipy}.

    Even in this straightforward setup, we note that the spectral method outperformed the FPC scheme in terms of runtime. However, it is worth pointing out that the FPC scheme demonstrates strong potential for parallelization, as highlighted by~\cite{Yoshikawa_2013}, who presented a detailed parallel implementation.

    \section{Numerical Tests}\label{sec: numerical test}
    In this section we perform numerical test for the VP equation for different initial condition, to verify the behavior of the integration scheme. All the simulations are done with $\rho^* = G = 1$. We have taken into consideration the evolution up to a single dynamical time $T = \sqrt{1 / G \rho^*} = 1$. As maximum velocity we use, 
    $V_{max} = 2 L / T$.

    \subsection{Timestep}\label{sec: timestep}
    The FPC scheme do not strictly necessitates for CLF condition, however for both scheme we have chosen to allow the maximum displacement for a single timestep to be a fraction $C = 1/2$ of the cell to ensure a more stable solution.
    \\
    We choose the starting timestep as in~\cite{Yoshikawa_2013}:
    \begin{equation}
        \label{eq: initial timestep LAX}
        \Delta t = C \cdot \min\left\{
            \Delta t_p = \frac{\Delta x}{V}; \, 
            \Delta t_v = \frac{\Delta v}{2 \max_i|{\partial_x \phi}_i|} \, ,
        \right\}
    \end{equation}
    where the factor $1/2$ comes from the fact that we do half timestep in the velocity integration. 
    The position timestep is constant throughout the all integration process and, while $\Delta t_v$ might change, because of the dependence on the force, does not happen in our simulations.
    %
    %
%
\begin{figure*}[t]
    \centering
    \includegraphics[width = 1.7\columnwidth]{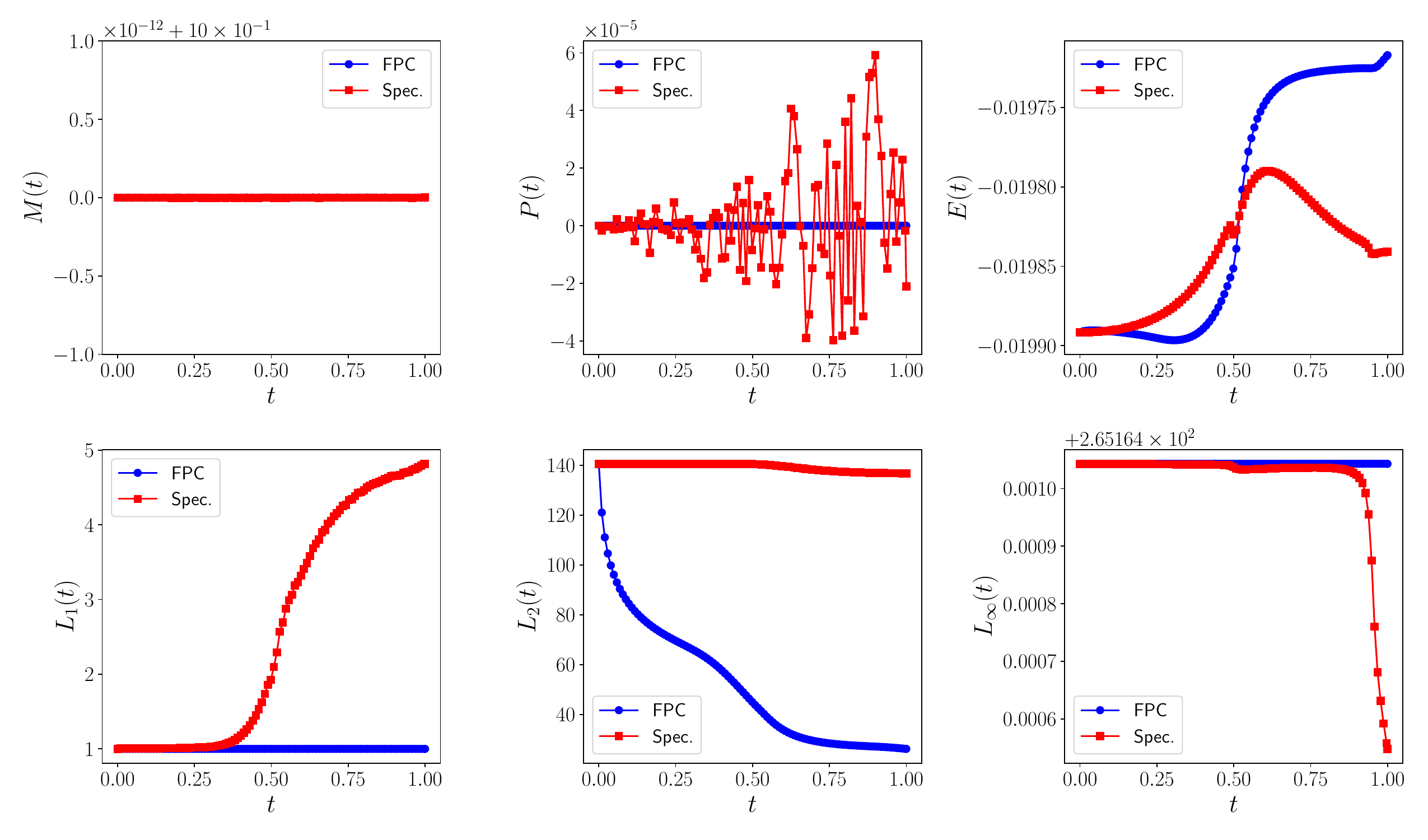}
    \caption{CDM smoothed initial condition with $\mathcal{R} = 0.004$: Evolution of physical observables (top row) and norm (bottom row) as a function of time in a simulation of a VP system with $N = 2^{10}$, $M = 2^{11} + 1$. Starting from the top-left corner we find, Mass $M(t)$, momentum $P(t)$ and energy $E(t)$; from bottom-left corner we find the $L_1$, $L_2$ and $L-\infty$ norm of the distribution function $f(x,v,t)$. }
    \label{fig: CDM smoothed observables}
\end{figure*}
    \subsection{Maxwellian initial condition}
    We will test the method proposed in Sec.~\ref{sec: numerical schemes} on smooth initial distribution, and verify the conservation laws aforementioned in Sec.~\ref{sec: vlasov for cdm}.
    
    A commonly used test for the accuracy of a chosen integration scheme for VP is the evolution of the Maxwellian distribution~\cite{comparison, Yoshikawa_2013, new-method-VP}
    \begin{equation}
        f(x,v, t) = \frac{\rho^*}{\sqrt{2\pi\sigma^2}} \exp \left( - \frac{v^2}{2\sigma^2} \right)
        \left[ 1 + A \, \sin\left(\frac{2\pi}{L}x - \frac{\pi}{2}\right) \right]
    \end{equation}
    with $A = 1/2$ and
    \begin{equation}\label{eq: sigma expression}
        \sigma^2 = \frac{4 \pi G \rho^*}{2 \pi / L} \mathcal{R}^2 \,.
    \end{equation}
    Adjusting the ratio $\mathcal{R} = k/k_j$ between the wave number $k = 2\pi / L$ and the jeans wave number $k_j$ we can move from linear to non-linear dynamics, specifically, by setting $\mathcal{R} = 1/2$, we study the case of non-linear Landau damping.
    
    Results are presented in Fig.~\ref{fig: Maxwellian observables}, where we tested both integration schemes on a grid composed by $N = 512$ by $M = 1024$ points, presenting the evolution of the physical observables that should be conserved by a VP system. 
    As it is shown, all quantities are preserved up to a given numerical precision, with the only side note concerning the fact that the spectral method conserves energy up to $10^{-4}$. We can thus state that all the results hint towards the fact that both integration schemes works correctly.
    
    
    \subsection{CDM Initial condition}
    \label{sec: CDM initial conditions}    
    Let us consider now the CDM case of a sinusoidal perturbation on the density:
        \begin{equation}
            \label{eq: initial rho}
            \rho(x,v,t=0) = 1 + A \, \sin\left(\frac{2\pi}{L}x - \frac{\pi}{2}\right) \, ,
        \end{equation}
        where everything is still, in other word the initial velocity distribution is given by a Dirac's delta $\delta(v-0)$. Thus our initial distribution function can be factorized in a velocity part and a spatial density.
        \begin{equation}
            \label{eq: initial distribution}
            f(x,v,t=0) = \delta(v, t=0) \, \rho(x,v,t=0) \, .
        \end{equation}

    Repeating the analysis performed for the Maxwellian initial condition, we simulate the evolution of a Vlasov–Poisson (VP) system on a $1024 \times 2049$ grid to ensure that both the spectral method and the FPC scheme conserve the physical quantities outlined in Sec.~\ref{sec: vlasov for cdm}. The results are shown in Fig.~\ref{fig: CDM delta observables}, where we observe that the FPC scheme conserves all quantities except for the $L_2$ norm and energy, with deviations limited to $10^{-4}$. As discussed in~\cite{comparison}, the inclusion of slope correctors in the FPC scheme accelerates the decay of the discrete $L_2$ norm. However, when oscillations caused by nonlinear effects are damped or smoothed through grid projection, the $L_2$ norm stabilizes.

\begin{figure*}[t]
    \includegraphics[width = 2\columnwidth]{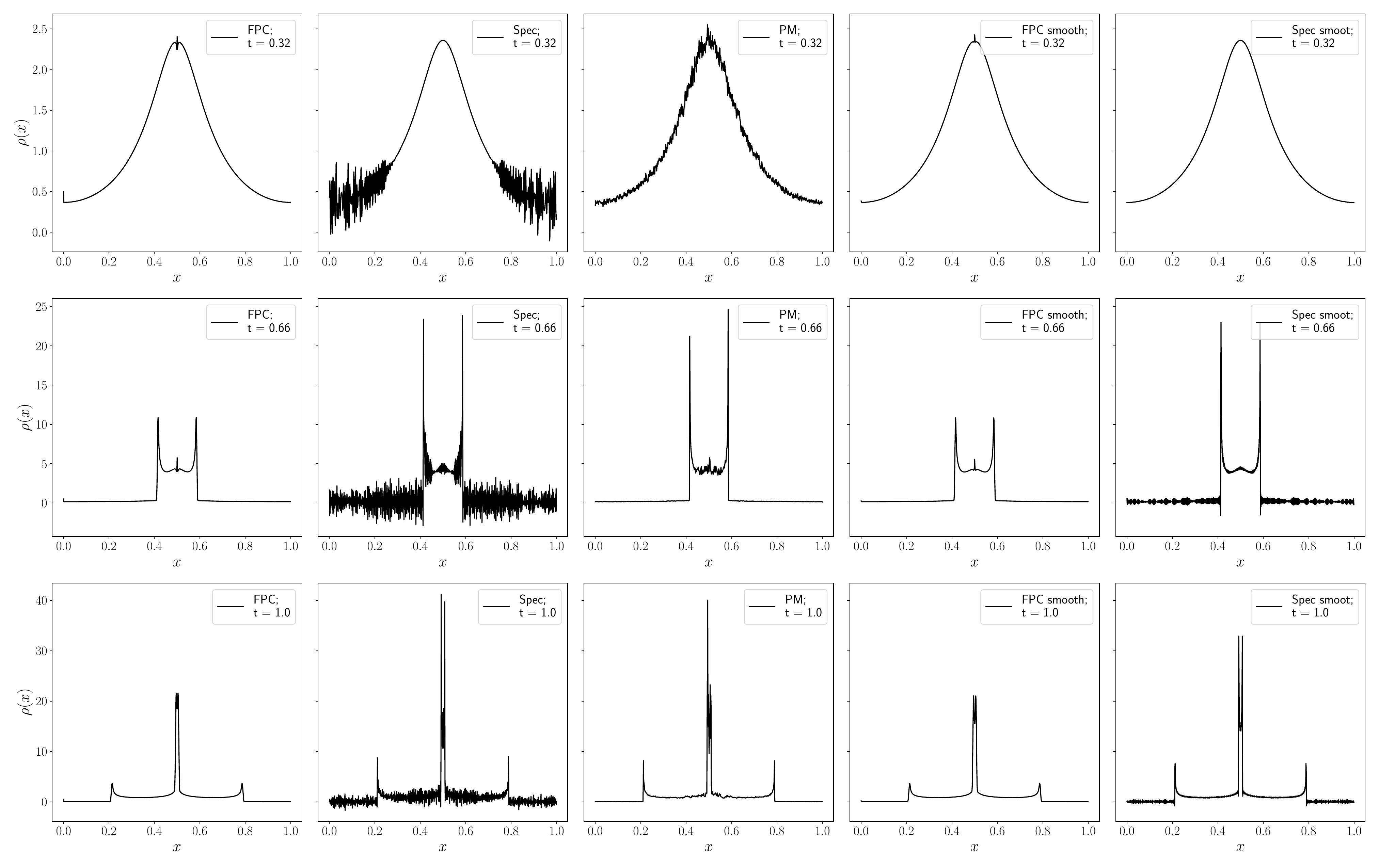}
    \caption{Comparison of densities $\rho(x)$ between VP results and a PM simulation (center column). The VP simulation are run on a $N \times M = 1024 \times 2049$ grid, while the PM simulation runs on $1024$ grid points and $10^6$ particles. The two column on left are CDM simulation done with the FPC and Spectral method respectively and discontinuous initial condition (Subsec.~\ref{sec: CDM initial conditions}), while the two on the right evolves the smoothed initial condition with $\mathcal{R} = 0.04$ (Subsec.~\ref{sec: smoothed condition}). The figure can be read row-wise or column-wise as on each row we compare the density $\rho(x)$ at the same time $t$ for different integration scheme, while on each column we can follow the evolution of the density for a given integration scheme. }
    \label{fig: denisty comp}
\end{figure*}

    This behavior is not observed in the spectral method. Without any limiter, spurious oscillations can cause the distribution function to become negative, resulting in violations of $L_1$ norm, momentum, and $L_\infty$ norm conservation.
    
    If the analysis is restricted to the conservation of the aforementioned quantities, one might conclude that the FPC scheme accurately reproduces the VP dynamics, while the spectral method does not. However, this conclusion is misleading. Fig.~\ref{fig: denisty comp} compares the densities obtained using the FPC scheme and the spectral method with those from a particle-mesh (PM) simulation, which serves as a reference. The first three columns reveal that the FPC scheme (first column) produces anomalous density peaks at $x = 0$ and $x = 1/2$. Furthermore, the density, $\rho$, appears smoother in comparison to both the spectral method (second column) and the PM simulation (center column). This smoothing arises from a combination of the stiff, discontinuous initial condition and the limiters applied in the FPC scheme.
    
    The discontinuity is most prominent along the velocity axis, where the function resembles $\delta(v)$. Analytical evolution would shift this function in accordance with Eq.~\eqref{eq: advection v}, yielding $\delta(v - F\Delta t)$. To compute these values numerically, interpolation is required. Fig.~\ref{fig: toy_model} illustrates how a delta function evolves under the FPC and spectral methods. Instead of propagating a delta function, $\delta(v - F\Delta t)$, as expected, the FPC scheme produces a smoothed, truncated version, which accounts for the smoothing observed in $\rho$.
    
    This issue does not arise in the spectral method, as the negative values retain information about the shifted delta in Fourier space. However, the spectral method introduces noise alongside the desired information, and spurious oscillations cannot be suppressed due to the absence of limiters.
    
    In addition to smoothing the distribution function, the above phenomenon causes two stationary density peaks to appear in the FPC results at $v = 0$ and $F = 0$. These two remain in fact stationary points during the whole evolution because the smoothed wavefunction prevents the potential (and consequently the force, $F$) from growing large enough to destabilize these points.


    \subsection{Smoothed initial condition}\label{sec: smoothed condition}
    If the previous consideration are true, the main obstacle is the Dirac's delta in the initial condition.
    We thus propose as alternative to the stiff discontinuous CDM initial condition, a smoother version, that however does not change the core of the dynamics. The Dirac's delta can be seen as an extreme case of a Gaussian distribution as $\sigma \to 0$. Thus we choose to recover the distribution function in Eq.~\eqref{eq: initial distribution} with small $\mathcal{R} = 0.004$ in order to obtain a softer version of the Dirac's delta, where instead of having one point different from zero we have three. In fact, from Eq.~\eqref{eq: sigma expression} we notice how choosing a small $\mathcal{R}$ corresponds to almost zero variance. This small numerical artifact provides a better staring point for both integration schemes at a small price: a very slight diffusion.

    We report the results obtained on a $1024 \times 2049$ simulation in Fig.~\ref{fig: CDM smoothed observables} and Fig.~\ref{fig: denisty comp}.

   By comparison with the results obtained in the previous section (Fig.~\ref{fig: CDM delta observables}), we immediately observe significant improvements in the spectral method.  
   The \(L_1\) norm transitions smoothly from 1 to a maximum of 5, in contrast to the abrupt discontinuity observed in Fig.~\ref{fig: CDM delta observables}. This behavior arises because smoothing of the initial conditions eliminates the initial presence of negative values in the distribution function. For the same reason, Fig.~\ref{fig: CDM smoothed observables} shows that momentum is conserved to a precision of \(10^{-4}\), compared to \(10^{-2}\) in the case of the discontinuous initial conditions.
    Additionally, the \(L_\infty\) norm is effectively conserved by the spectral method when the initial conditions are smoothed, as the relative variation is \(\mathcal{O}(10^{-4}/10^2) = \mathcal{O}(10^{-6})\). Finally, the \(L_2\) norm exhibits an approximate variation of order unity, corresponding to a relative variation of \(\mathcal{O}(10^{-2})\). This is significantly smaller than the relative variation of \(\mathcal{O}(1)\) observed for the discontinuous case (Fig.~\ref{fig: CDM delta observables}).

    This improvement arises because the spectral method is highly sensitive to steep discontinuities, lacking mechanisms to regulate spurious oscillations. 
    However, despite the improvements introduced by the spectral method once the stiffness issue is resolved, no clear-cut winner between spectral and FPC methods emerges from examining the physical quantities shown in Fig.~\ref{fig: CDM smoothed observables}. Therefore, we turn our attention to the behavior of the density function.
    
    In Fig.~\ref{fig: denisty comp}, we can clearly see how smoothing has benefited both the FPC and spectral schemes. For the spectral method (rightmost column), oscillations are significantly damped, and while negative densities persist, they are comparable to minor numerical noise. The peak heights align almost perfectly with the PM results. A closer examination reveals a slight diffusion effect due to the smoothed initial condition, as the function lacks the sharp edges observed in its counterpart with discontinuous initial conditions (second column).
    
    Regarding the FPC scheme (second column from the right), we observe only minor changes. The density is slightly less smooth, but the most notable improvement is that the peak at $x = 0$ has almost disappeared. This supports our hypothesis that the primary issue lies in the integration of a steep, discontinuous distribution function. Smoothing the initial condition has effectively eliminated one of the peaked structures.

    Even though the spectral method exhibits negative values, which are non-physical, the overall physical observables are accurately reproduced when the smoothed initial condition is evolved using the spectral method.
    %
    %
    %
    \section{Summary and Conclusions}

    In this paper, we addressed the challenge of integrating a Vlasov–Poisson (VP) system for cold dark matter (CDM) simulations, where the initial condition exhibits strong discontinuities. In such cases, commonly used integration schemes face significant difficulties and fail to accurately reproduce the desired density. We identified the root cause of this issue as the steep discontinuity in the initial condition.
    
    To overcome this problem, we proposed a method that introduces a small softening in the initial distribution, ensuring that the final results remain unaltered.
    
    We conducted tests using both a spectral integrator and the FPC scheme on a $1024 \times 2049$ grid, for both the steep initial condition and its adjusted version. 
    Our findings indicate that the smoothed spectral method is the best choice for integrating discontinuous initial condition such as CDM.

    The results can be summarized as follows:
    \begin{itemize}
        \item \textbf{First test:} For a smooth Maxwellian initial condition, both the spectral integrator and the FPC scheme were shown to conserve all the quantities expected for a VP system.
        \item \textbf{Second test:} For a VP system simulating a CDM distribution with a discontinuous initial condition:
        \begin{itemize}
            \item The FPC scheme conserved mass, momentum, $L_1$ norm, $L_\infty$ norm and energy (up to $10^{-4}$), but failed to accurately reproduce the density and distribution function. Furthermore, the $L_2$ was not preserved.
            \item The spectral method, while failing to conserve most observables, correctly reproduced the shape of the density and distribution function. However, non-physical negative values were observed.
        \end{itemize}
        \item \textbf{Proposed solution:} By introducing a smoother initial condition, we ensured the dynamics of the system were preserved. Under these circumstances, the spectral method accurately reproduced the density and distribution function while conserving mass, momentum (up to $10^{-5}$), energy (up to $10^{-4}$), and the $L_2$ and $L_\infty$ norms. The only drawback was the appearance of negative values in the distribution function and density, which we consider comparable to numerical noise.
    \end{itemize}
        
    In conclusion, we recommend addressing discontinuous VP problems by applying a very small smoothing to the initial condition, so that a spectral method can be used to solve the VP system of equations, with appropriate care to numerical noise. 
    The smoothing of the initial distribution prevents the development of unphysical behaviour in the numerical solution, without qualitatively altering the nature of the problem. Future developments of this analysis could involve considering approach based on anti-aliasing or filamentation filters, as presented in\cite{klimas} that, applied on top on the smoothing, could lead to a significant noise reduction.


\section*{Acknowledgment}
We want to thank Erik Sonnendr\"ucker for useful discussions and insights.

\bibliographystyle{IEEEtran}
\bibliography{ref}

\begin{thebibliography}{10}
\providecommand{\url}[1]{#1}
\csname url@samestyle\endcsname
\providecommand{\newblock}{\relax}
\providecommand{\bibinfo}[2]{#2}
\providecommand{\BIBentrySTDinterwordspacing}{\spaceskip=0pt\relax}
\providecommand{\BIBentryALTinterwordstretchfactor}{4}
\providecommand{\BIBentryALTinterwordspacing}{\spaceskip=\fontdimen2\font plus
\BIBentryALTinterwordstretchfactor\fontdimen3\font minus \fontdimen4\font\relax}
\providecommand{\BIBforeignlanguage}[2]{{%
\expandafter\ifx\csname l@#1\endcsname\relax
\typeout{** WARNING: IEEEtran.bst: No hyphenation pattern has been}%
\typeout{** loaded for the language `#1'. Using the pattern for}%
\typeout{** the default language instead.}%
\else
\language=\csname l@#1\endcsname
\fi
#2}}
\providecommand{\BIBdecl}{\relax}
\BIBdecl

\bibitem{Planck2020}
{Planck Collaboration}, N.~{Aghanim}, and {other}, ``{Planck 2018 results. VI. Cosmological parameters},'' \emph{Astronomy and Astrophysics}, vol. 641, p.~A6, Sep. 2020.

\bibitem{Weinberg2013}
D.~H. {Weinberg}, M.~J. {Mortonson}, D.~J. {Eisenstein}, C.~{Hirata}, A.~G. {Riess}, and E.~{Rozo}, ``{Observational probes of cosmic acceleration},'' \emph{Physical Reports}, vol. 530, no.~2, pp. 87--255, Sep. 2013.

\bibitem{Mocz_towards}
\BIBentryALTinterwordspacing
P.~Mocz and A.~Szasz, ``Toward cosmological simulations of dark matter on quantum computers,'' \emph{The Astrophysical Journal}, vol. 910, no.~1, p.~29, mar 2021. [Online]. Available: \url{https://dx.doi.org/10.3847/1538-4357/abe6ac}
\BIBentrySTDinterwordspacing

\bibitem{Cappelli}
\BIBentryALTinterwordspacing
L.~Cappelli, F.~Tacchino, G.~Murante, S.~Borgani, and I.~Tavernelli, ``From vlasov-poisson to schr\"odinger-poisson: Dark matter simulation with a quantum variational time evolution algorithm,'' \emph{Phys. Rev. Res.}, vol.~6, p. 013282, Mar 2024. [Online]. Available: \url{https://link.aps.org/doi/10.1103/PhysRevResearch.6.013282}
\BIBentrySTDinterwordspacing

\bibitem{Springel2016}
V.~{Springel}, ``{High Performance Computing and Numerical Modelling},'' in \emph{Saas-Fee Advanced Course}, ser. Saas-Fee Advanced Course, Y.~{Revaz}, P.~{Jablonka}, R.~{Teyssier}, and L.~{Mayer}, Eds., vol.~43, Jan. 2016, p. 251.

\bibitem{Yoshikawa_2013}
\BIBentryALTinterwordspacing
K.~Yoshikawa, N.~Yoshida, and M.~Umemura, ``Direct integration of the collisionless boltzmann equation in six-dimensional phase space: Self-gravitating systems,'' \emph{The Astrophysical Journal}, vol. 762, no.~2, p. 116, dec 2012. [Online]. Available: \url{https://dx.doi.org/10.1088/0004-637X/762/2/116}
\BIBentrySTDinterwordspacing

\bibitem{widrow-kaiser}
L.~M. {Widrow} and N.~{Kaiser}, ``{Using the Schroedinger Equation to Simulate Collisionless Matter},'' \emph{The Astrophysical Journal Letters}, vol. 416, p. L71, Oct. 1993.

\bibitem{mocz_schrodinger-poissonvlasov-poisson_2018}
\BIBentryALTinterwordspacing
P.~Mocz, L.~Lancaster, A.~Fialkov, F.~Becerra, and P.-H. Chavanis, ``Schrödinger-{Poisson}–{Vlasov}-{Poisson} correspondence,'' \emph{Physical Review D}, vol.~97, no.~8, p. 083519, Apr. 2018. [Online]. Available: \url{https://link.aps.org/doi/10.1103/PhysRevD.97.083519}
\BIBentrySTDinterwordspacing

\bibitem{cheng-knorr}
C.~Z. Cheng and G.~Knorr, ``The integration of the vlasov equation in configuration space,'' \emph{Journal of Computational Physics}, 1976.

\bibitem{sonnendrucker}
\BIBentryALTinterwordspacing
E.~Sonnendrücker, J.~Roche, P.~Bertrand, and A.~Ghizzo, ``The semi-lagrangian method for the numerical resolution of the vlasov equation,'' \emph{Journal of Computational Physics}, vol. 149, no.~2, pp. 201--220, 1999. [Online]. Available: \url{https://www.sciencedirect.com/science/article/pii/S0021999198961484}
\BIBentrySTDinterwordspacing

\bibitem{filbet-sonnendrucker-fpc}
\BIBentryALTinterwordspacing
F.~Filbet, E.~Sonnendrücker, and P.~Bertrand, ``Conservative numerical schemes for the vlasov equation,'' \emph{Journal of Computational Physics}, vol. 172, no.~1, pp. 166--187, 2001. [Online]. Available: \url{https://www.sciencedirect.com/science/article/pii/S0021999101968184}
\BIBentrySTDinterwordspacing

\bibitem{klimas}
\BIBentryALTinterwordspacing
A.~Klimas and W.~Farrell, ``A splitting algorithm for vlasov simulation with filamentation filtration,'' \emph{Journal of Computational Physics}, vol. 110, no.~1, pp. 150--163, 1994. [Online]. Available: \url{https://www.sciencedirect.com/science/article/pii/S0021999184710114}
\BIBentrySTDinterwordspacing

\bibitem{oliver-hahn}
\BIBentryALTinterwordspacing
O.~Hahn and R.~E. Angulo, ``An adaptively refined phase–space element method for cosmological simulations and collisionless dynamics,'' \emph{Monthly Notices of the Royal Astronomical Society}, vol. 455, no.~1, pp. 1115--1133, 11 2015. [Online]. Available: \url{https://doi.org/10.1093/mnras/stv2304}
\BIBentrySTDinterwordspacing

\bibitem{comparison}
\BIBentryALTinterwordspacing
F.~Filbet and E.~Sonnendrücker, ``Comparison of eulerian vlasov solvers,'' \emph{Computer Physics Communications}, vol. 150, no.~3, pp. 247--266, 2003. [Online]. Available: \url{https://www.sciencedirect.com/science/article/pii/S001046550200694X}
\BIBentrySTDinterwordspacing

\bibitem{new-method-VP}
\BIBentryALTinterwordspacing
D.~Yi and S.~Bu, ``A mass conservative scheme for solving the vlasov–poisson equation using characteristic curve,'' \emph{Journal of Computational and Applied Mathematics}, vol. 324, pp. 1--16, 2017. [Online]. Available: \url{https://www.sciencedirect.com/science/article/pii/S0377042717301899}
\BIBentrySTDinterwordspacing

\bibitem{hockney-eastwood}
R.~Hockney and J.~Eastwood, \emph{Computer Simulation Using Particles}.\hskip 1em plus 0.5em minus 0.4em\relax CRC Press., 1988.

\end{thebibliography}

\end{document}